\documentclass[conference]{IEEEtran}
\usepackage{graphicx}
\usepackage{algorithm}
\usepackage{rotating}
\usepackage{colortbl}
\usepackage{amsmath}
\usepackage{multirow}
\usepackage{balance}
\usepackage{enumerate}
\usepackage{amssymb}
\usepackage{url}
\usepackage[numbers]{natbib}
\ifCLASSINFOpdf
\else
\fi
\begin{document}
%
\title{Towards Modeling the Interaction of Spatial-Associative Neural Network Representations for Multisensory Perception}

\author{\IEEEauthorblockN{German I. Parisi$^1$, Jonathan Tong$^2$, Pablo Barros$^1$, Brigitte R\"oder$^2$, Stefan Wermter$^1$}
\IEEEauthorblockA{$^1$Knowledge Technology, Department of Informatics, Universit\"at Hamburg, Germany\\
$^2$Biological Psychology and Neuropsychology, Universit\"at Hamburg, Germany\\ \\
\textit{Workshop on Computational Models for Crossmodal Learning, IEEE ICDL-EPIROB 2017, Lisbon, Portugal}
}
}
\maketitle

\begin{abstract}
Our daily perceptual experience is driven by different neural mechanisms that yield multisensory interaction as the interplay between exogenous stimuli and endogenous expectations.
While the interaction of multisensory cues according to their spatiotemporal properties and the formation of multisensory feature-based representations have been widely studied, the interaction of spatial-associative neural representations has received considerably less attention.
In this paper, we propose a neural network architecture that models the interaction of spatial-associative representations to perform causal inference of audiovisual stimuli.
We investigate the spatial alignment of exogenous audiovisual stimuli modulated by associative congruence.
In the spatial layer, topographically arranged networks account for the interaction of audiovisual input in terms of population codes.
In the associative layer, congruent audiovisual representations are obtained via the experience-driven development of feature-based associations.
Levels of congruency are obtained as a by-product of the neurodynamics of self-organizing networks, where the amount of neural activation triggered by the input can be expressed via a nonlinear distance function.
Our novel proposal is that activity-driven levels of congruency can be used as top-down modulatory projections to spatially distributed representations of sensory input, e.g. semantically related audiovisual pairs will yield a higher level of integration than unrelated pairs.
Furthermore, levels of neural response in unimodal layers may be seen as sensory reliability for the dynamic weighting of crossmodal cues.
We describe a series of planned experiments to validate our model in the tasks of multisensory interaction on the basis of semantic congruence and unimodal cue reliability.
\end{abstract}


%
\IEEEpeerreviewmaketitle

\section{Introduction}

Perception comprises multiple sensory modalities to allow for a robust and efficient interaction with the environment~\cite{Stein93}.
In humans and non-human mammals, multisensory perception is mediated by a rich set of neural mechanisms providing the means to process multisensory stimuli on the basis of their spatiotemporal alignment~\cite{Macaluso2004} and semantic congruence~\cite{Laurienti2004}.
Such biological mechanisms have inspired neurocomputational approaches aimed at effectively modeling multisensory perception and behavior also in the presence of unreliable input and crossmodal conflicts.
However, computational architectures often account for the processing of either spatially, temporally, or semantically related stimuli.

Successful computational approaches have been proposed that integrate (or segregate) simplified audiovisual stimuli (typically light blobs and beeps) according to their spatial and temporal properties~(e.g. \cite{Ursino17}), i.e. maximum integration is obtained for spatially aligned, co-occurring audiovisual pairs; whereas stimuli which are far away in space and time should be segregated.
More complex behavior arises when subjects are exposed to spatially discrepant or asynchronous audiovisual pairs within small spatial or temporal windows, respectively, yielding biased responses such as in the spatial ventriloquism effect (auditory stimulus shifted towards the position of the synchronous visual one,~\cite{Jack73}) and its temporal variant (asynchronous sound modulating the perceived visual onset).
Computational approaches typically address this problem with the implementation of a causal inference model~\cite{Kording17}, where the integration of an auditory and a visual stimulus is dictated by the probability of the two stimuli to be generated by a common source.

In addition to spatiotemporal cues, we can relate multisensory information based on semantic congruence~\cite{Laurienti2004}.
This congruence can be inferred from prior knowledge and expectations about the properties of specific crossmodal events.
In this context, a number of computational models have been proposed that address the multisensory binding of feature-based representations such as the image of a dog and the sound of a dog's bark.
Specific associations can be obtained through the exposure of a learning architecture (e.g. a neural network) to a set of different audiovisual pairs~\cite{Vavrecka2013}\cite{Ursino2015b}.
However, these models generally focus on feature-based processing and rule out the important role of spatial information.
In complex environments (for instance, consider seeing multiple dogs instead of only one or seeing both dogs and cats), spatial and temporal information allow for solving the correspondence problem, that is, which sound to bind to which animal.
Behavioral and neurophysiological studies on crossmodal tasks evidence the strong interplay of spatial-associative representations that together contribute to the development of a robust percept including situations with conflicts and degraded sensory information.
Although brain areas involved in the processing of spatial and feature-based representations have been widely studied both anatomically and functionally, the mechanisms of interaction of spatial-associative representations are barely understood.
Accordingly, there is a lack of computational approaches that parsimoniously model these two aspects (space and semantics) into one unified learning framework.
The development of such models is interesting not only from the perspective of a better understanding of the neural network dynamics underlying multisensory perception, but also from a practical perspective with artificial systems operating in natural (crossmodal) environments~(see \cite{Ursino14} for a review).

\begin{figure}[t!] 
\center{\includegraphics[width=0.95\linewidth]{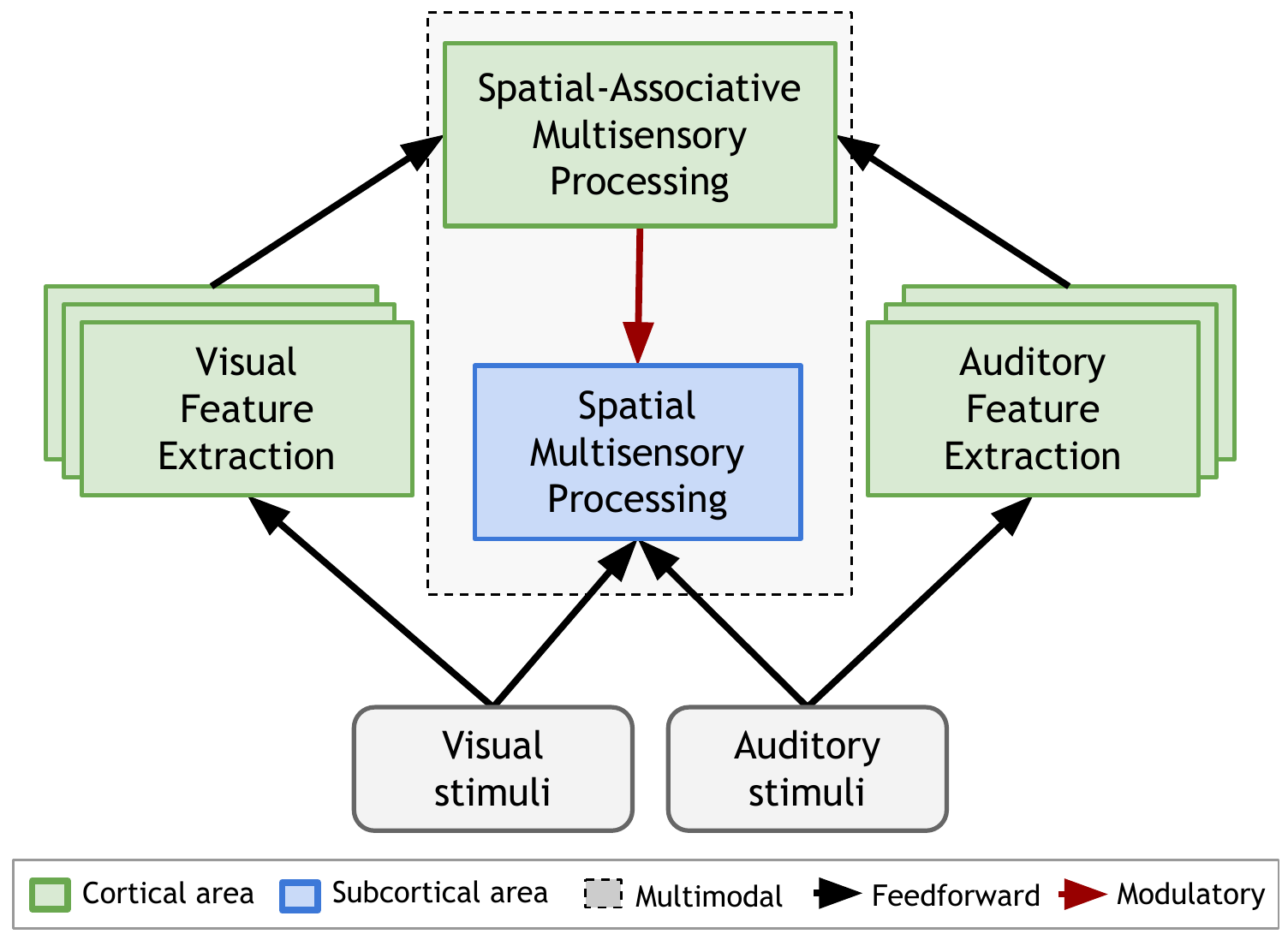}}
\caption{Conceptual illustration of our learning architecture for multisensory perception.}
\label{fig:overview}
\end{figure}

In this paper, we suggest a neural network architecture for learning spatial-associative representations of audiovisual stimuli.
A conceptual view of our architecture is shown in Fig.~\ref{fig:overview}, where visual and auditory stimuli are processed in parallel with different levels of complexity.

This architecture is inspired by multisensory processing in the brain, where a subcortical area -the superior colliculus~(SC)- processes audiovisual patterns on the basis of their spatiotemporal correspondence~\cite{Stein93}.
Concurrently, stimuli are processed by the cortical areas -the visual and the auditory cortex- with information accumulated over increasingly large spatiotemporal windows.
These hierarchies of neural detectors extract unimodal features with increasing representational power via feedforward and recurrent connectivity patterns.\footnote{For simplicity, we assume that the visual and auditory hierarchies account for the processing of unisensory representations. However, there is evidence for strong crossmodal interactions, e.g. with auditory representations modulating visual perception.}
Unisensory representations converge towards higher-order brain areas that learn and respond to multisensory representations~\cite{Malacuso2005}: with the ventral occipital and the superior temporal sulcus (STS) involved in location-independent feature-based processing, lateral and dorsal occipital areas being activated by spatiotemporally concurring stimuli, and the right inferior parietal lobule activated for synchronous, spatially discrepant events.

Complex networks of neurons interact with the aim to provide a joint multisensory percept and to trigger robust behavioral responses, with expectations driven by internal representations modulating the perception of exogenous stimuli.
It has been suggested that the spatial-associative modulation is mediated by top-down connections from these high order areas to the SC, with the latter using information from the former to solve the causal inference problem.
However, the exact neural mechanisms underlying this process are to be fully investigated.

In the next section, we describe and motivate the use of neural network modules for modeling the interaction of spatial-associative representations, with these spatial and associative neural representations being learned by exposing the networks to a set of training audiovisual patterns.
Although the modeling of specific brain areas is out of the scope of our work, we integrate a number of neural processing principles that are well accounted for by the neurodynamics of our architecture.
Finally, we describe a set of experiments to validate our model in the context of a robotic scenario where multisensory perception modulated by semantic congruence is necessary to trigger adequate behavioral responses.


\section{Proposed Model}

\subsection{Overview}

Our architecture is composed of a set of self-organizing neural networks which learn multisensory representations through the exposure to congruent audiovisual input pairs.
Both spatial and feature-based representations are obtained via the unsupervised training of multiple network layers.

The subcortical module~(Fig.~\ref{fig:overview}, blue box) is responsible for processing audiovisual pairs on the basis of their spatial and temporal alignment (similar to the SC).
In this case, neurons are topologically aligned and each neuron encodes for a specific position of space (details in Section II.B).

The cortical modules are responsible for extracting auditory and visual features with increasing complexity, with the spatial-associative multisensory module learning high-level representations of audiovisual events.
Levels of congruency obtained as a function of neural activity in this layer are used to modulate spatial representations in the subcortical module.


Connectivity patterns in the brain are known to develop from an experience-driven learning process.
Accordingly, our ability to integrate crossmodal stimuli and solve crossmodal conflicts is progressively acquired and fine-tuned through the exposure to multimodal events~\cite{Wallace04}.
Self-organizing networks learn prototype representations of the input space without supervision, yielding topology-preserving feature maps as the result of the Hebbian-like update rule of neural weights and connectivity patterns.
Therefore, the neurodynamics of self-organizing systems provide an interesting framework for the development of neural representations and the emergence of inter-layer connectivity patterns.

\subsection{Subcortical Module}

\begin{figure} 
\center{\includegraphics[width=\linewidth]{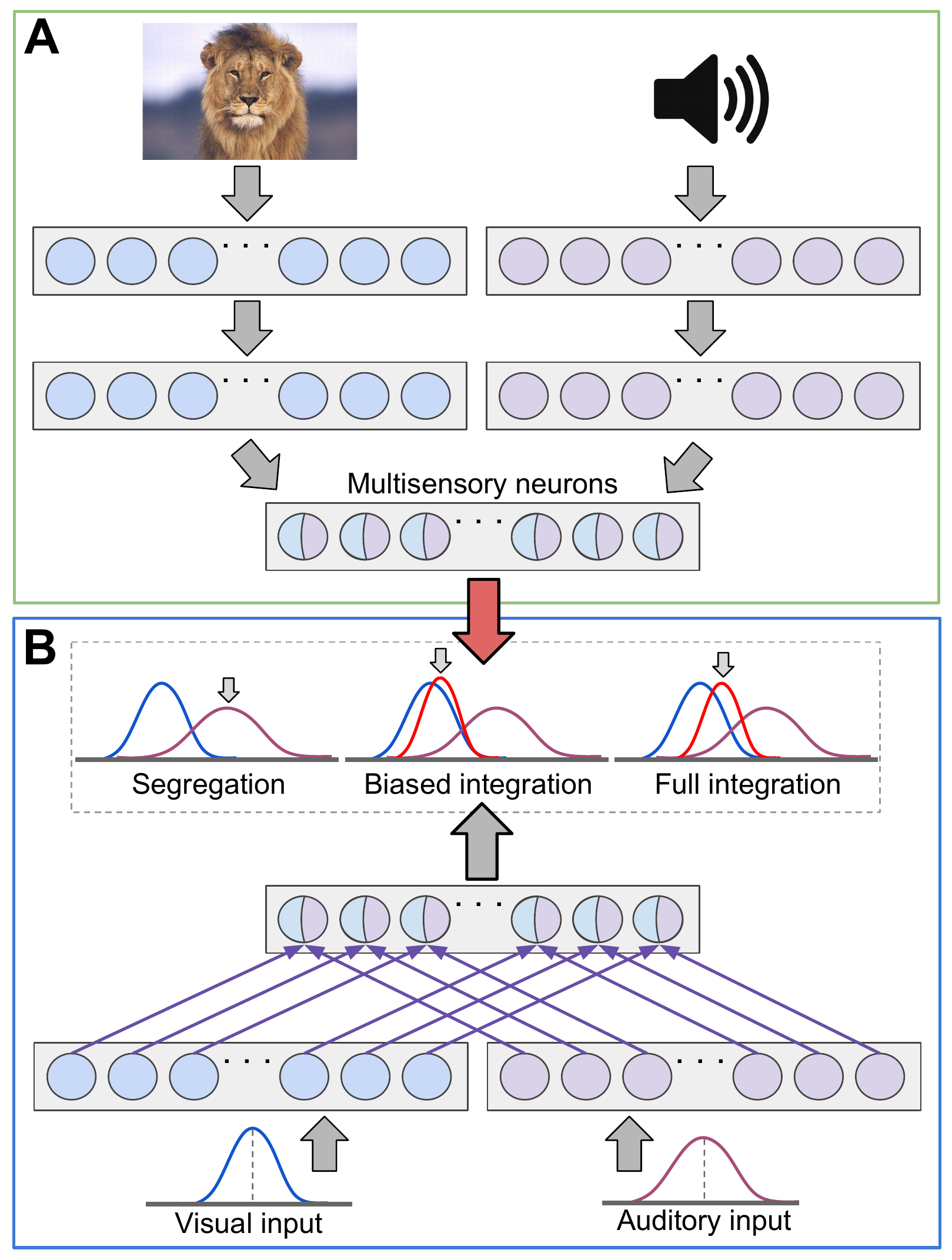}}
\caption{Cortico-collicular architecture for acquiring multisensory integration using self-organizing neural layers. Top-down feedback from high-order neural representations (Box A) is used to modulate multisensory interaction on the basis of spatial alignment (Box B).}
\label{fig:background}
\end{figure}

The subcortical module consists of two upstream layers of $N$ visual and $N$ auditory neurons and a downstream layer with $N$ multisensory neurons~(Fig.~\ref{fig:background}, Box~B).
This architecture is based on the two-layer architecture extended with a third layer for the causal inference problem~\cite{Ursino17}.
Neurons are topologically aligned and each neuron codes for a specific position of space (e.g. with $N=180$).
The visual and auditory input are represented by Gaussian functions resembling spatially localized external stimuli filtered by the receptive fields of unisensory neurons.
The model assumes that the auditory and visual area are spatially organized, with the spatial resolution of auditory neurons being smaller than the spatial resolution of visual input.

Neurons in the two unisensory layers are reciprocally connected through inter-area excitatory synapses~(Fig.~\ref{fig:background}, violet lines) so that each neuron receives input only from the neuron of the other modality at the same spatial position.
These connections modulate the influence of one modality over the other.
Therefore, each output of a neuron within a modality $m$ (with $m=v$ or $m=a$) processes input as the sum of the external input, the intra-area lateral input, and the inter-layer crossmodal input.

Pre-trained intra-layer connections define neural receptive fields, whereas crossmodal connections are trained via the exposure to unimodal and crossmodal input.
The learning of connectivity patterns is carried out via Hebbian training rules for synaptic potentiation.
Hebbian-like development of inter-layer connections accounts for the ventriloquism effect, where the perception of auditory stimulus is shifted in the direction of the visual one provided that the spatial discrepancy between the two stimuli is smaller than 20-25 degrees~\cite{Magosso12}.
Conversely, when the spatial discrepancy is higher, the effect of the integration of the two stimuli is negligible.
This behavior is consistent with a Bayesian estimator that sub-optimally computes the prior and likelihood probabilities for inferring the position of crossmodal stimuli~\cite{Ursino17}.
 
\subsection{Feature-based Multimodal Associations}

The learning and the recognition of meaningful visual and auditory patterns is implemented as a hierarchy of self-organizing networks that tune internal representations to process features with an increasing degree of complexity and abstraction~\cite{Parisi2017}.
This framework provides a mechanism to develop experience-driven associations of complex audiovisual patterns, i.e. videos and sounds~(Fig.~\ref{fig:background}, Box~A).

For each modality $m$, a self-organizing hierarchy will compute a location-invariant, feature-based representation of the input $\textbf{x}^m$ (with prototype neural representations denoted as $\textbf{w}^m$).
Multisensory neural representations $\textbf{w}^M$ of the input $\textbf{x}^{M}$ are obtained via the training of the high-level network with tuples of the form $<\textbf{w}^v,\textbf{w}^a>$.
The amount of neural activation $\Psi^M$ triggered by $\textbf{x}^M$ can be expressed via a nonlinear function $\Psi^M = \exp(-r) \in (0,1]$, where $r=\Vert \textbf{x}^M - \textbf{w}^M \Vert$ is the reconstruction error defined by the discrepancy between the input and its neural representation (i.e. maximum activation for $r=0$).
In this setting, $\textbf{w}^M$ will better match the crossmodal input if both $\textbf{w}^v$ and $\textbf{w}^a$ represent well the unimodal input.

We propose that levels of crossmodal congruency can be obtained as a by-product of the neurodynamics of self-organizing networks, with $\Psi^M$ representing the level of neural response to crossmodal events and yielding the following relation: 
\begin{equation}
\Psi^M \propto \phi(\Psi^v,\Psi^a),
\end{equation}
where $\phi$ denotes a nonlinear transformation from unisensory responses to the response of a multisensory association developed through experience-driven unsupervised learning.
This neural activity behavior qualitatively resembles neural responses found in the STS and the middle temporal gyrus (MTG)~\cite{Beauchamp2004}, with multisensory neurons exhibiting the highest activation for congruent audiovisual patterns.
STS/MTG neurons fire for incongruent patterns as well but exhibiting a decreased response.

\subsection{Spatial-Associative Modulation}

Causal Bayesian inference considers the two hypotheses that a given crossmodal stimulus is either caused by a common cause or by independent causes~\cite{Kording17}, from which it is then possible to derive optimal predictions of multisensory integration.
However, the prior probability that there is a single cause versus
two causes ($p_{common}$) is a parameter that does not consider any semantic relation, thus stimuli are integrated entirely on the basis of their spatial relation.

It is expected that semantically related audiovisual pairs will yield a higher level of integration with respect to incongruent ones~\cite{Habets17}.
Therefore, activity-driven levels of congruency in the associative layer can be used as top-down modulatory projections to the spatial layer with the aim to bias the integration or segregation of crossmodal stimuli~(Fig.~\ref{fig:background}, red arrow).
More specifically, the parameter $p_{common}$ can be dynamically computed as a function of $\Psi^M$.

Multisensory interaction is a dynamic process that takes into account the reliability of each sensory cue, with human observers linearly combining available cues by weighting them in proportion to their reliability~\cite{Fetsch2012}.
For instance, while it is often the case that visual cues spatially dominate auditory ones due to a higher spatial resolution, strongly blurred visual stimuli yield the opposite effect.
In this setting, we speculate that the characteristic of dynamically reweighting crossmodal cues on the basis of their reliability may be obtained as a by-product of the neurodynamics of the associative layer.
Levels of neural activity in the unisensory layers may provide the means to compute the reliability of each modality, i.e. estimating the weights as a function of $\Psi^m$.

\section{Planned Experiments}

We plan to conduct a series of experiments to evaluate our model for the following tasks: i) integrating crossmodal input in the spatial layer; ii) modulating spatial integration on the basis of activity-driven levels of semantic congruence; and iii) dynamic weighting of crossmodal cues according to unimodal reliability.
For this purpose, during the training session we will expose our neural network architecture to a set of congruent audiovisual stimuli (videos and sounds of animals).
The training will be carried out in an unsupervised fashion, with multisensory associations developing according to the co-occurrence audiovisual pairs.

After the training, the exposure to learned associations of congruent audiovisual patterns should yield an adequately strong neural response (stronger responses for incongruent patterns).
We will conduct post-training experiments in three different conditions: i) audiovisual pairs (containing one animal) with or without spatial/congruent discrepancy; ii)~audiovisual pairs (containing multiple animals in the visual modality) and one auditory stimulus with or without spatial discrepancy; and iii)~the previous two conditions varying the level of visual degradation (e.g. blurriness) for testing mechanisms of dynamic cue reliability.

A behavioral study may be conducted that reproduces these experimental conditions to human subjects in order to trigger human-like responses in artificial systems exposed to crossmodal events.
Although the exact neural mechanisms underlying optimal multisensory interaction remain to be unveiled, neurocomputational models may provide the means to study the interaction of complex neural networks for acquiring multisensory perception.

\section*{Acknowledgment}

\small{This research was supported by National Natural Science Foundation of China (NSFC) and the German Research Foundation (DFG) under project Transregio Crossmodal Learning (TRR 169).}



%

\balance
\bibliographystyle{ieeetr}
\bibliography{cmlbiblio2}

%
%

\end{document}